\documentclass[aps,twocolumn,floatfix]{revtex4}

\usepackage{graphicx,amsmath}

\begin{document}

\title{ The diffuse Galactic $\gamma$-rays from dark matter annihilation }

\author{Xiao-Jun Bi$^{1,2}$}
\email{bixj@mail.ihep.ac.cn}
\author{Juan Zhang$^{1}$}
\author{Qiang Yuan$^{1}$}
\author{Jian-Li Zhang$^{1}$}
\author{HongSheng Zhao$^{3}$}
\affiliation{$^{1}$Key laboratory of particle astrophysics, IHEP, 
Chinese Academy of Sciences, Beijing 100049, P. R. China, \\
$^{2}$ Center for High Energy Physics,
Peking University, Beijing 100871, P.R. China\\ 
$^{3}$ University of St Andrews, School of Physics and Astronomy,
KY16 9SS, Fife, UK}

\begin{abstract}

The diffuse Galactic $\gamma$-rays from EGRET observation shows excesses
above 1~GeV in comparison with the expectations from conventional
Galactic cosmic ray (CR) propagation model. In the work we try to solve the
``GeV excess'' problem by dark matter (DM) annihilation in the frame of
supersymmetry (SUSY). Compared with  previous works, there are 
three aspects improved in this work: first, the direction-independent 
``boost factor'' for diffuse $\gamma$-rays from dark matter annihilation 
(DMA) is naturally reproduced by taking the DM substructures into account; 
second, there is no need for renormalization of the diffuse $\gamma$-ray 
background produced by CRs; last but not the least, in this work our new propagation model 
can give consistent results of both diffuse $\gamma$-rays and antiprotons, 
by directly adding the signals from DMA to the diffuse $\gamma$-ray background.
This is a self-consistent model among several possible
scenarios at present, and can be tested or optimized by the forthcoming
experiments such as GLAST, PAMELA and AMS02.

\end{abstract}

\maketitle

The diffuse Galactic $\gamma$-rays are produced via interaction
of CRs with the interstellar medium and radiation field. However,
the spectrum of the diffuse $\gamma$ rays measured by EGRET shows 
an excess above $1$~GeV \cite{hunter} in comparison with the 
prediction based on the conventional CR model, whose nucleus and 
electron spectra are consistent with the locally observed data.
The discrepancy may indicate large-scale proton or electron 
spectrum, which determines the diffuse $\gamma$-rays, different 
than the local measured one, or the existence of exotic sources 
of diffuse continuum $\gamma$-ray emission.

A harder nucleon spectrum with power-law index of $-2.4\sim2.5$ 
has been proposed in Ref.~\cite{gral} to solve the ``GeV excess'' 
problem. However, it has been pointed out that such a hard nucleon 
spectrum will overproduce secondary antiprotons and positrons 
\cite{moska}, which has effectively been excluded by recently high 
energy $\bar{p}/p$ ratio measurements \cite{beach}. A hard electron 
spectrum is studied in Ref.~\cite{pp} while this hypothesis also 
suffers difficulties, e.g. it produced too many $\gamma$-rays at 
higher energies and couldn't be compatible with the local electron 
spectrum \cite{optimize}. For the ``optimized model'' 
in \cite{optimize} both the proton and electron injection spectra 
are ``fine-tuned'' and their intensities are renormalized to 
explain the EGRET diffuse $\gamma$ spectra. However, it may be 
not easy for the proton spectrum to fluctuate significantly and 
to be different from other heavy nuclei, as introduced in \cite{optimize}.

It is shown that the observed peak of the diffuse $\gamma$ spectrum
at low galactic latitudes, where the dominant contribution is from 
pion decay, is at higher energies than the $\pi^0$ decay peak. 
Further the conventional model with reacceleration is known \cite{moska02} 
to produce less antiprotons at $\sim 2$~GeV than the measurement at BESS 
\cite{Orito00} by a factor of about $ 2$. Positron data also show some 
``excess'' at higher energies \cite{heat}. These discrepancies may 
all indicate a contribution from ``exotic'' sources, e.g. DMA \cite{anni}.

de Boer et al. \cite{boer} pointed out that the ``GeV excess'' could be 
explained by the long-awaited signal of DMA from the Galactic halo. 
By fitting both the background spectrum from cosmic nucleon collisions 
and the signal spectrum from neutralino, the lightest supersymmetric 
particle, annihilation they found the EGRET data could be well explained 
in all directions. From the spatial distribution of the diffuse 
$\gamma$-ray emission they constructed the DM profile,  with two rings 
supplemented on the smooth halo. A direction independent ``boost factor'' 
to the signal flux usually at the order of $100$ is necessary to explain 
the $\gamma$-ray excess. Another factor between $1/2-2$ for the 
background flux is also needed to account for the spectra at different 
directions. However, de Boer's model with ring profiles and a large 
boost factor will lead to possible conflict with the antiproton flux, 
as shown by Bergstr\"om et al. \cite{bergs}.

Based on the model-fitting by de Boer et al. \cite{boer} and Strong's work \cite{optimize},
we try to explain the diffuse $\gamma$-ray spectrum 
in this work by {\em directly} calculating the background
and DMA fluxes and to overcome their shortcomings at the same time.
By adjusting the propagation parameters
we try to give consistent descriptions to the measured spectra without any
arbitrary normalization of the background contribution. We calculate
the DMA in the frame of the minimal supersymmetric extension of the
standard model (MSSM). After taking into account the enhancement by
the existence of subhalos \cite{minihalo} we do not need the
``boost factor'' any more. Furthermore in our propagation model,
we found the antiproton flux is in agreement with the measurements.
The crucial point is that the enhancement by subhalos is spatial
dependent in the Galactic halo, not ``universal'' as the previous works
adopted. So the enhancement of $\gamma$-ray is different from that of
antiproton flux, because the whole halo will contribute to the diffuse
$\gamma$-ray intensity, while only antiprotons produced within the diffusion region will contribute
to the observed flux. It is found that the same scenario with large
boost by subhalos can be used to explain the positron excess \cite{yuan}.

The fluxes of DMA products are determined by two independent factors.
The first factor is related to the annihilation cross section and
determined by particle physics of DM, while the other one is connected with
the spatial distribution of DM and determined by astrophysics \cite{anni}.
We use the package DarkSUSY \cite{darksusy} to calculate the particle
physical factor of DMA. Scanning the parameter space of MSSM we find
the $\gamma$-ray spectrum with $m_\chi=40\sim50$ GeV can fit the EGRET
data well. The branching ratios between neutralino annihilation into
${\bar p}$ and $\gamma$-rays are also calculated for different MSSM
parameters and are found to be $1/20\sim 1/10$ in a wide mass range.
We chose a $m_\chi =48.8$ GeV model which predicts
$\Omega h^2 = 0.09$ and $\frac{Br(\chi\chi\to {\bar p})}
{Br(\chi\chi\to\gamma)}\approx 0.055$ for energies above the threshold
$E_{th} = 0.5$ GeV.  The second factor determining the annihilation 
fluxes is defined as $\Phi^{astro}=\int\frac{\rho^2}{D^2} \text{d}V$ with $D$ 
the distance to the source of $\gamma$-ray production, $\rho$ the
density profile of DM and $V$ the volume of annihilation taking place.  
When we consider the contribution from subhalos, the factor
is given by the number integral along a direction $(\theta,\phi)$,
$\Phi_{sub}=\int_{l.o.s.} \Phi^{astro}\text{d}N_{sub}(\theta,\phi) $.
We use the simulation result of the subhalo distribution with
mass $m_{sub}$ at the radius $r$ \cite{bi,diemand} as $N_{sub}(m_{sub},r)
= N_0 \left( \frac{m_{sub}}{M_v} \right)^{-1.9} \left(
1+\left( \frac{r}{r_H} \right)^2\right)^{-1}$,
where $M_v \approx 1.0\times 10^{12} M_{\odot}$ is the mass of the Galaxy, 
$r_H=0.14 r_v\approx 36$ kpc ($r_v\approx260$ kpc is the virial radius
of the Galaxy halo) is the core radius for the distribution of subhalos, 
$r$ is the distance to the Galactic center (GC) and $N_0$ is the 
normalization factor. The minimal subhalos can be as light as 
$10^{-6} M_{\odot}$ as shown by the recent simulation conducted by 
Diemand et al. \cite{minihalo}, while the maximal mass of substructures 
is taken to be $0.01 M_v$ \cite{bi}. The tidal effects are taken into 
account under the ``tidal approximation'' \cite{bi} so that the 
subhalos are disrupted near the GC. The total signal flux comes 
from the annihilation in the subhalos and the smooth component.

\begin{figure}
\resizebox{8.5cm}{6cm}{\includegraphics{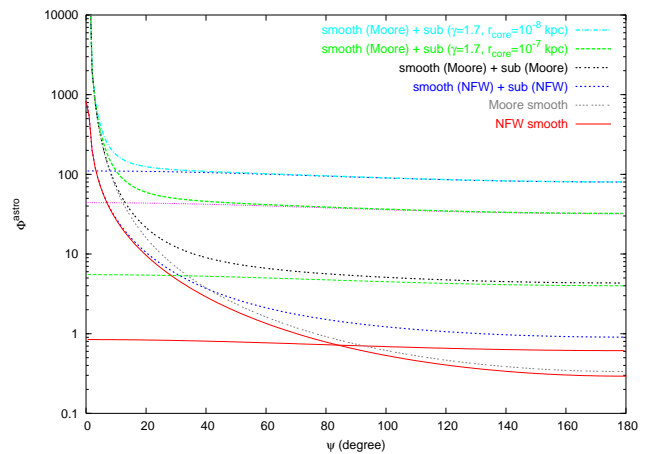}}
\caption{ \label{density}
The astrophysics factor $\Phi^{astro}$ (in unit of 
(GeV/cm$^3$)$^2$~kpc~Sr$^{-1}$) from different directions. The almost 
horizontal lines correspond to the contributions from subhalos only.}
\end{figure}
                                                                                
The DM density profile within each subhalo is taken as
the NFW \cite{nfw97}, Moore \cite{m99} or a cuspier form \cite{zhao} as
$\rho = \frac{\rho_s}{(r/r_s)^\gamma (
1+r/r_s)^{3-\gamma}}$ with $\gamma = 1.7$. The last form
is favored by the simulation conducted by Reed et al. \cite{reed},
which shows that $\gamma = 1.4 - 0.08\log(M/M_*)$ increases for
smaller subhalos. We take $\gamma = 1.7$ for the whole range
of subhalo masses as a simple approximation. The small halos
with large $\gamma \approx 1.5\sim2$ are also found by Diemand et al.
\cite{minihalo}. To determine the profile parameters, we also need to know
the concentration $c_v$ as a function of halo mass. Here we adopted
the semi-analytic model of Bullock et al. \cite{bullock}, which describes
$c_v$ as a function of virial mass and redshift. We adopt the mean
$c_v-m_{sub}$ relation at redshift zero (see also Fig.~1 of Ref.~\cite{yuan}).
The scale radius is then determined as $r^{nfw}_s=r_v/c_v$,
$r^{moore}_s=r^{nfw}_s/0.63$ or
$r^\gamma_s=r^{nfw}_s/(2-\gamma)$. Another factor determining the
$\gamma$-ray flux is the core radius, $r_{\text{core}}$,
within which the DM density should be kept constant due to the balance
between the annihilation rate and the infalling rate of DM particles
\cite{berezinsky}. The core radius $r_{\text{core}}$ is
approximately in the range $10^{-8} \sim 10^{-7}$ kpc for the $\gamma = 1.7$
profile and $10^{-9} \sim 10^{-8}$ kpc for the Moore profile.
In Fig.~\ref{density} we show the factor $\Phi^{astro}$ from the
smooth component, the subhalos and the total contribution as a function
of the direction to the GC. The $\Phi^{astro}$ from subhalos is almost
isotropic to different directions, this is because the DM distribution
is almost spherical symmetric and the Sun is near the GC.
We can see that the largest enhancement for $\gamma = 1.7$ subhalos
at large angles can reach 2 orders of magnitude and depends on the value
of $r_{\text{core}}$, while for the Moore profile the enhancement is about one
order of magnitude and for NFW profile only about $3$ times larger.
The $\Phi^{astro}$ for Moore and NFW profiles is not sensitive to
$r_{\text{core}}$ \cite{bi}. We also notice that near the GC there is no enhancement.
This is actually a very important difference from the model given by de Boer
\cite{boer} where the ``boost factor'' is universal.
Given the factor $\Phi^{astro}$ and the SUSY model we can predict the
$\gamma$-ray flux by neutralino annihilation.

\begin{figure*}
\resizebox{8.5cm}{6.5cm}{\includegraphics{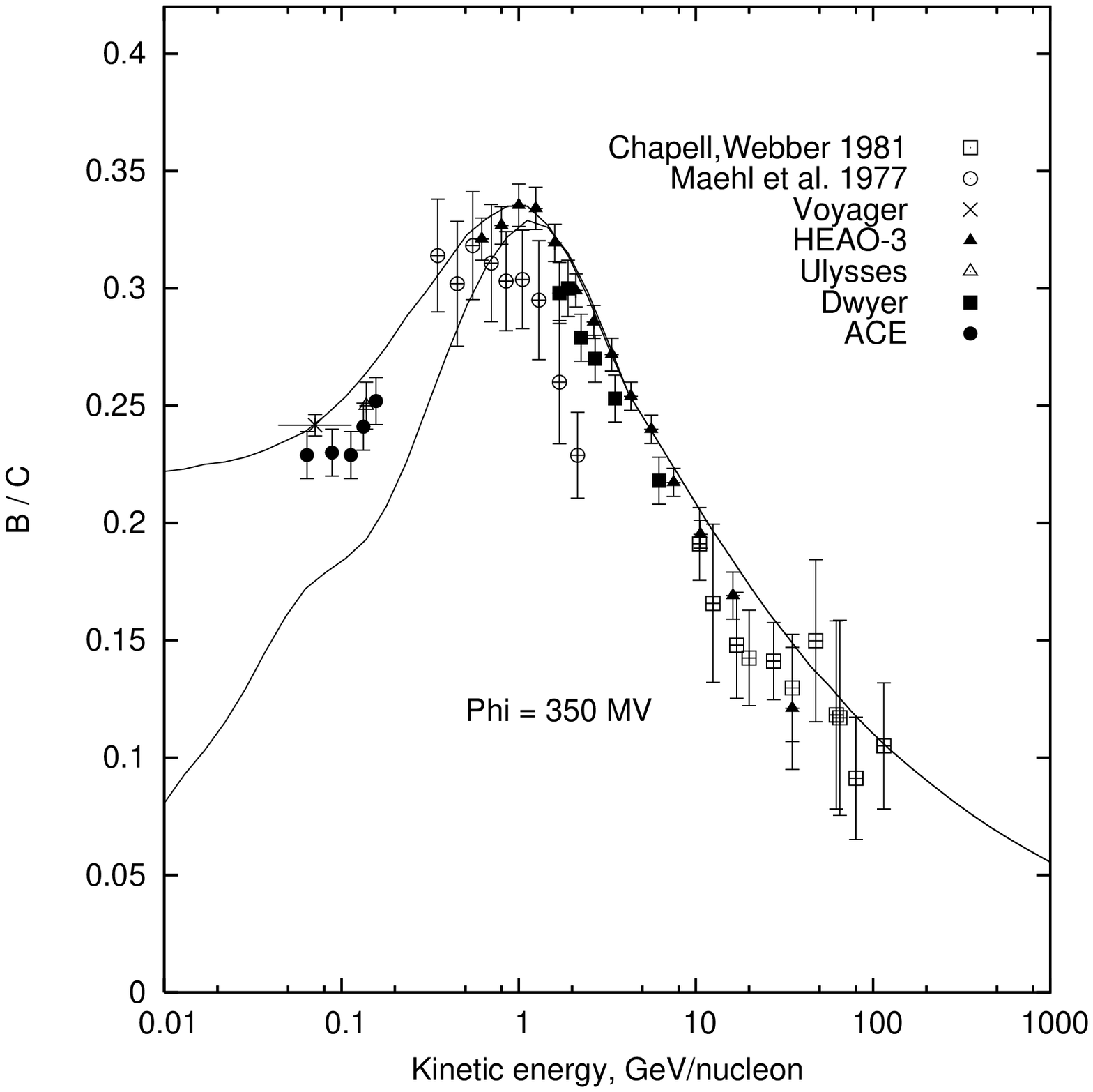}}
\resizebox{8.5cm}{6.5cm}{\includegraphics{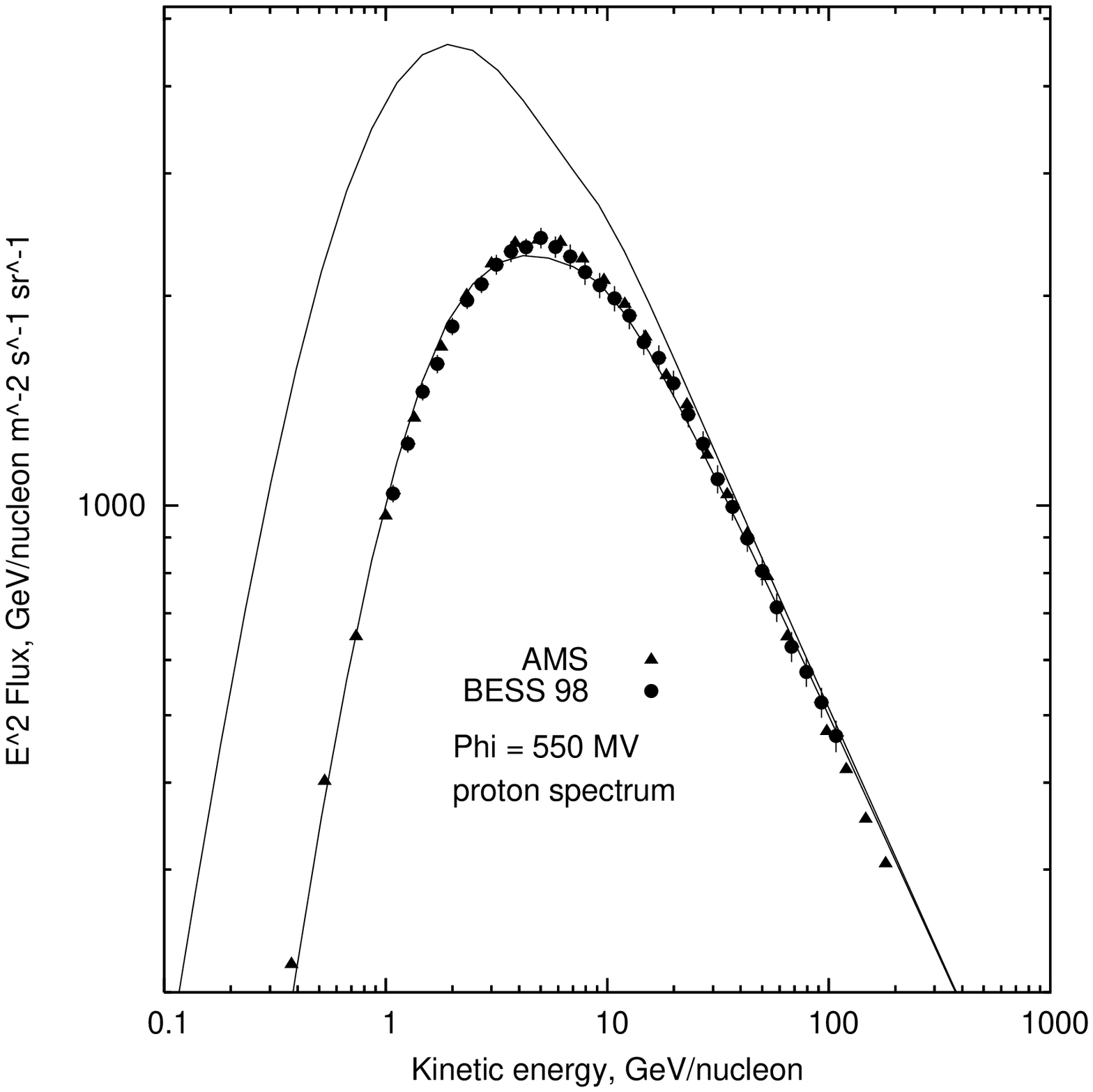}}
\caption{\label{prop}
B/C and proton spectrum in the present model. 
Lower curve for B/C is LIS, upper is the modulated,
while the lower curve for proton spectrum is the modulated 
and upper is LIS.
For the experimental data, see \cite{optimize}.
}
\end{figure*}

We now turn to the calculation of the background 
diffuse $\gamma$-ray emission,
which consists of several components:
the neutral pion decay produced by energetic interactions of nuclei 
with interstellar gas, emission by electrons inverse Compton scattering 
off the interstellar radiation field, 
the bremsstrahlung of electrons in interstellar medium,
and the extragalactic background.
We calculate the background diffuse $\gamma$-rays using the package
GALPROP \cite{sm} which uses the realistic distributions 
for the interstellar gas and radiation fields and solves the diffusion
equations numerically.

We have paid extreme effort to calculate the background so that
we can give good description to the EGRET data after adding the DMA component.
The injection spectra of protons and heavier nuclei are assumed 
to have the same power-law form in rigidity. We include the nuclei
up to $Z=28$ and relevant isotopes. 
For propagation, we use the diffusion
reacceleration model \cite{mo02}. 
The diffusion halo height of the propagation is taken as
$z_h=1.5$ kpc, which is different from 4 kpc as adopted in \cite{optimize,
boer}. A smaller $z_h$ can effectively lower the $\bar{p}$ flux
since it is only $\bar{p}$ from DMA in the diffusion region that
can contribute to the flux observed on the Earth.
The propagation parameters have been tuned to fit the B/C ratio and
the local proton (and electron) spectra, as shown in Fig.~\ref{prop}. 
A major uncertainty in the models of diffuse Galactic $\gamma$-ray
emission is the distribution of molecular hydrogen for the derivation of
H$_2$ density from the CO data is problematic \cite{sm04}.
For example, the scaling factor $X_{\text{CO}}$ from COBE/DIRBE 
studies by Sodroski et al. \cite{sodroski}
is about $2-5$ times greater than the value given by Boselli et al. 
\cite{boselli} in different Galactocentric radius based on the
measurement of Galactic metallicity gradient and the
inverse dependence of $X_{\text{CO}}$ on metallicity, which is normalized to 
the $\gamma$-ray data \cite{sm04}. 
An analysis of EGRET diffuse $\gamma$-ray emission yields 
a constant $X_{\text{CO}} = (1.9\pm 0.2)\times 10^{20}$cm$^{-2}/$(K km s$^{-1}$) for
$E_\gamma = 0.1 - 10$~GeV \cite{sm96}. Observations of particular
local clouds yield lower values $X_{\text{CO}} = 0.9-1.65 
\times 10^{20}$ cm$^{-2}$/(K km s$^{-1}$).
Since the fit to the EGRET data for $E_\gamma = 0.1 - 10$~GeV
in \cite{sm96} assumes
only the background contributions, we expect they give larger $X_{\text{CO}}$
than the case with new components, such as the consideration here.
We find a smaller $X_{\text{CO}}=0.6\sim 1.0 \times 10^{20}$ molecules 
cm$^{-2}/$(K km s$^{-1}$) can give much better fit to the EGRET data
below $1$~GeV. We take $X_{\text{CO}}$ a constant independent of the 
radius $R$. As shown in Ref.~\cite{sm04} the
simple form is compensated by an appropriate form of the CR
sources.
We have taken the radial distribution of CR sources in the form
of $(r/{r_o})^\alpha e^{-\beta (r-{r_o})/{r_o}} $  
with $\alpha=1.35$, $\beta=2.7$, ${r_o}=8.5$\,kpc,  
and limiting the sources within $r_{max}=15$ kpc,
which are adjusted to best describe the diffuse $\gamma$-ray spectrum.

\begin{figure*}
\includegraphics[scale=0.35]{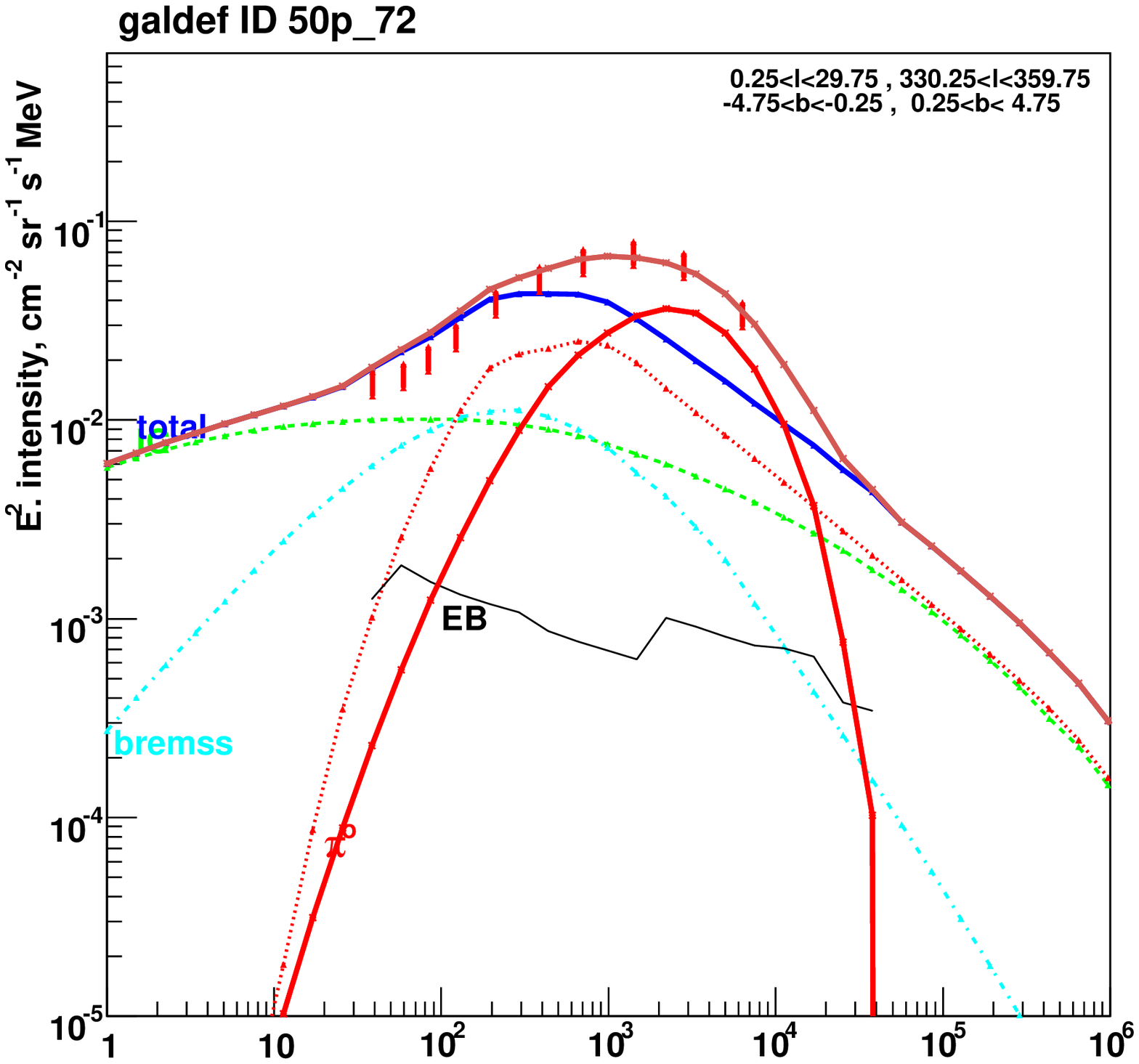}
\includegraphics[scale=0.35]{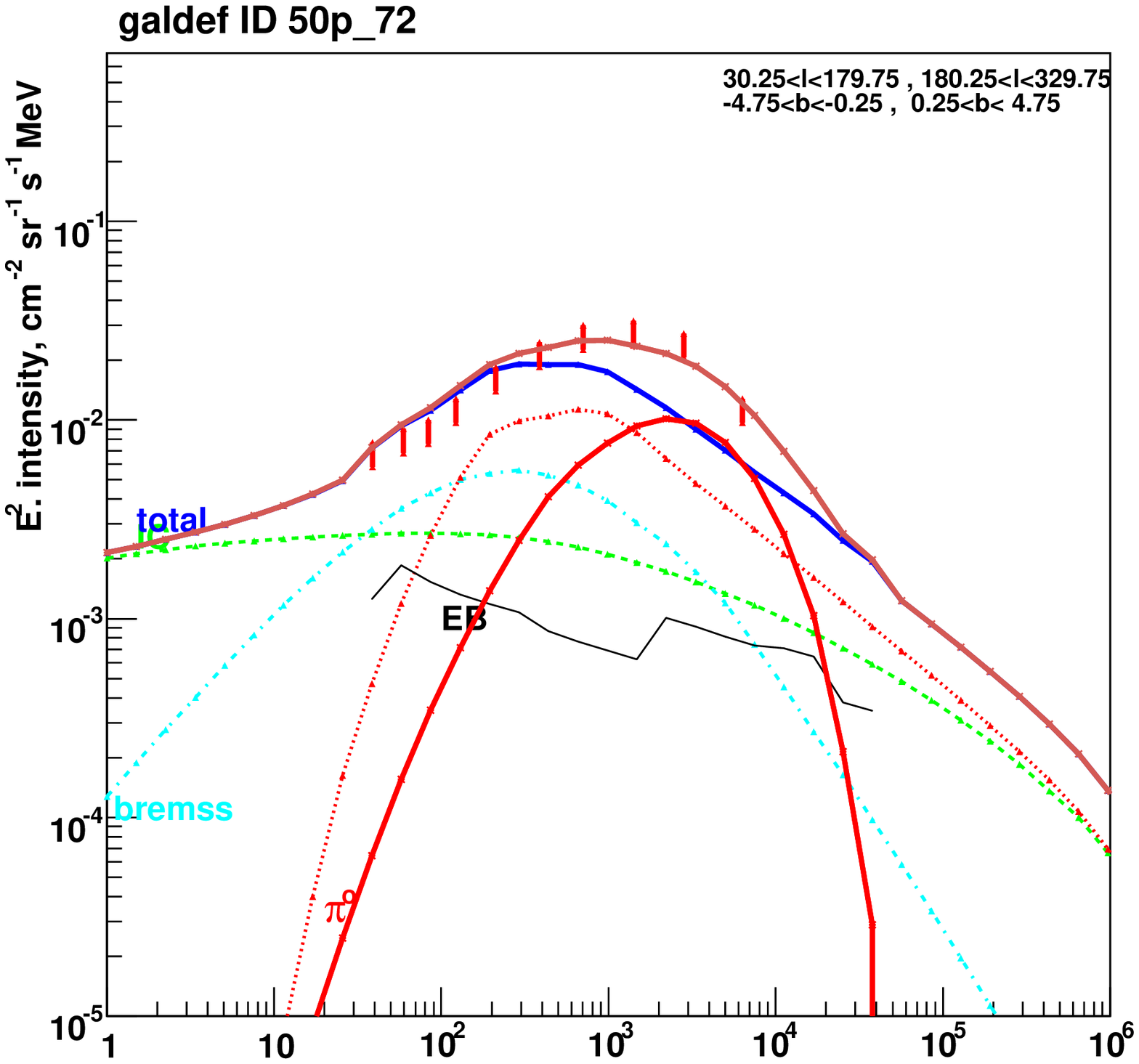}
\includegraphics[scale=0.35]{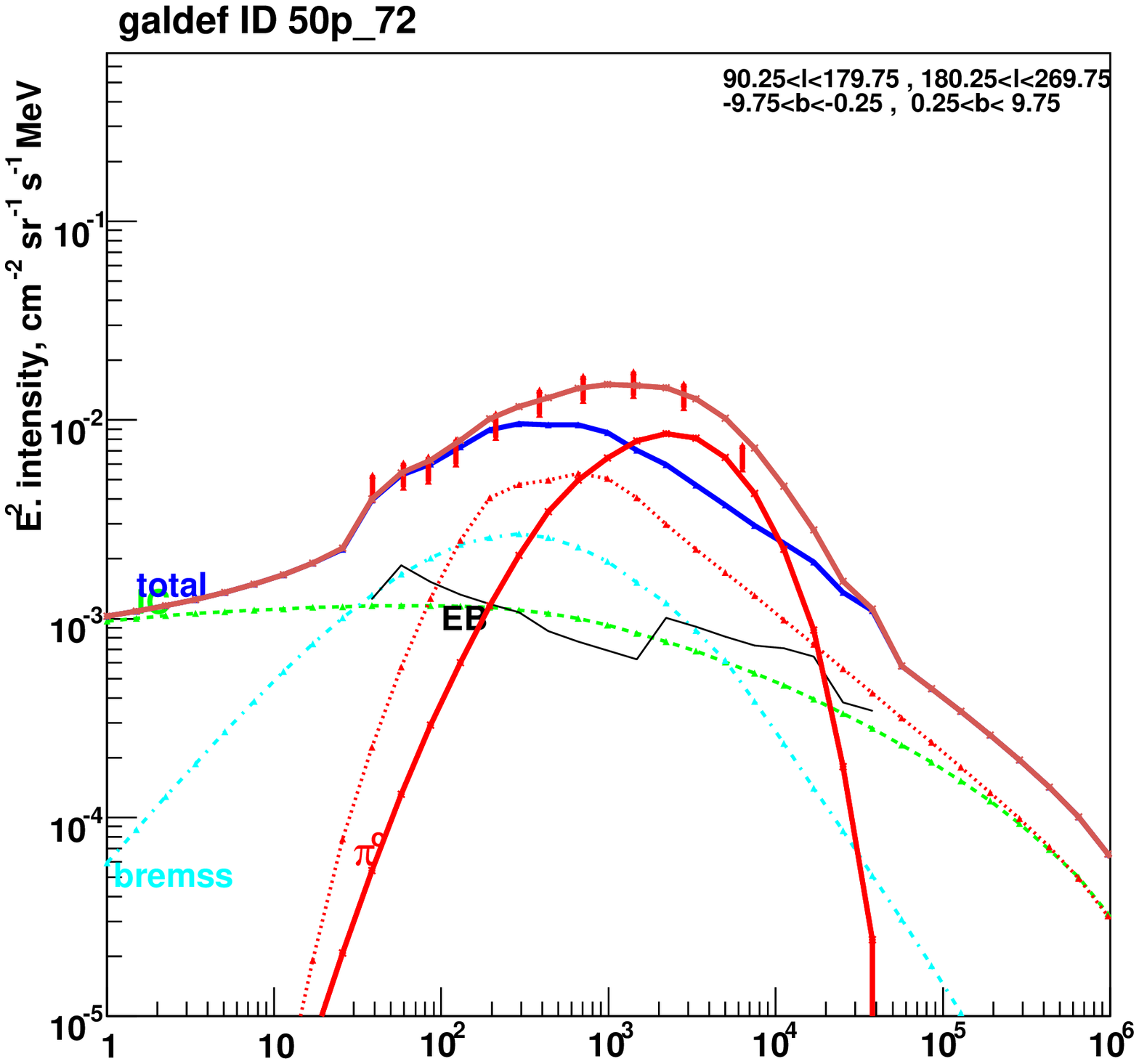}
\includegraphics[scale=0.35]{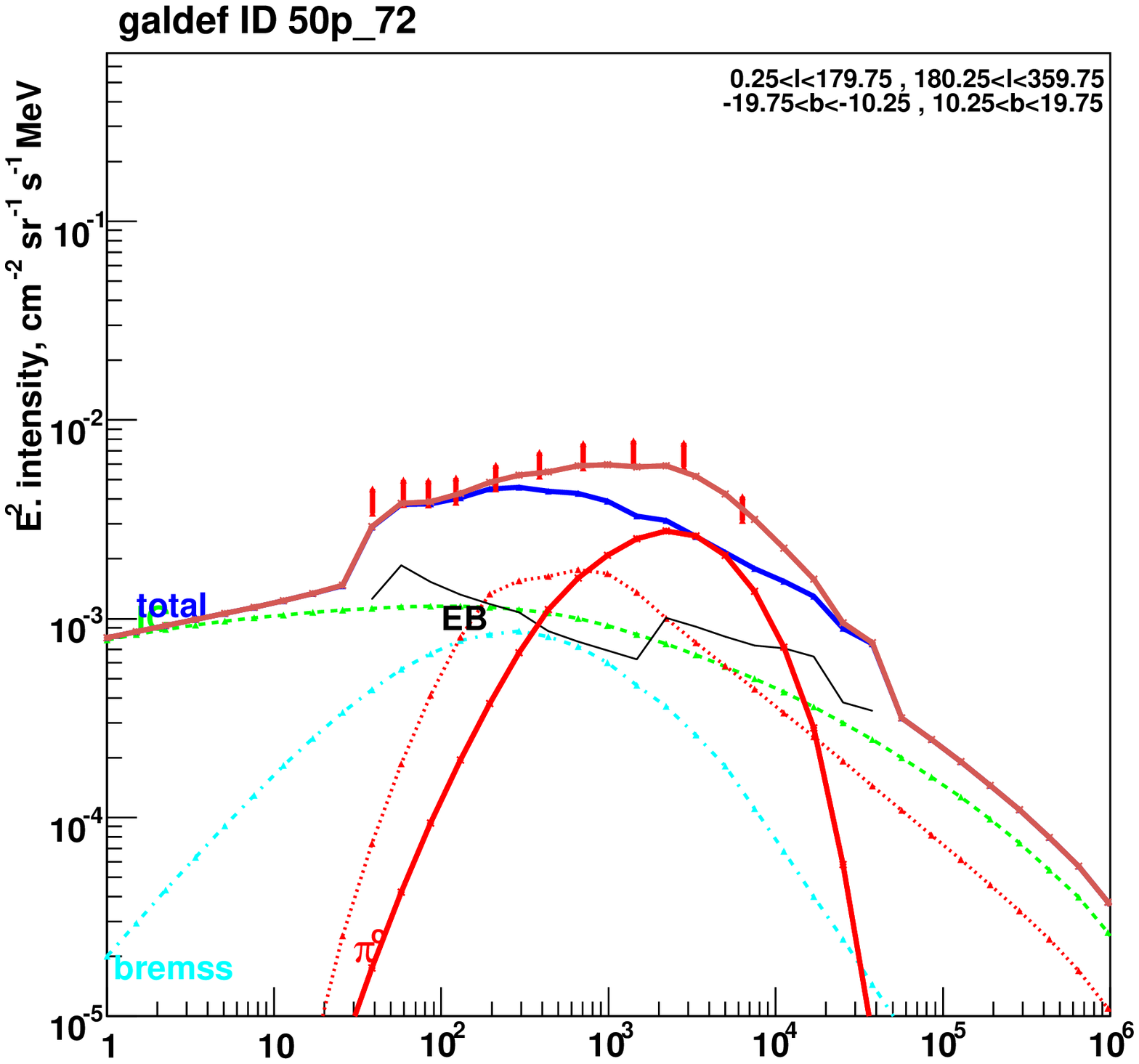}
\includegraphics[scale=0.35]{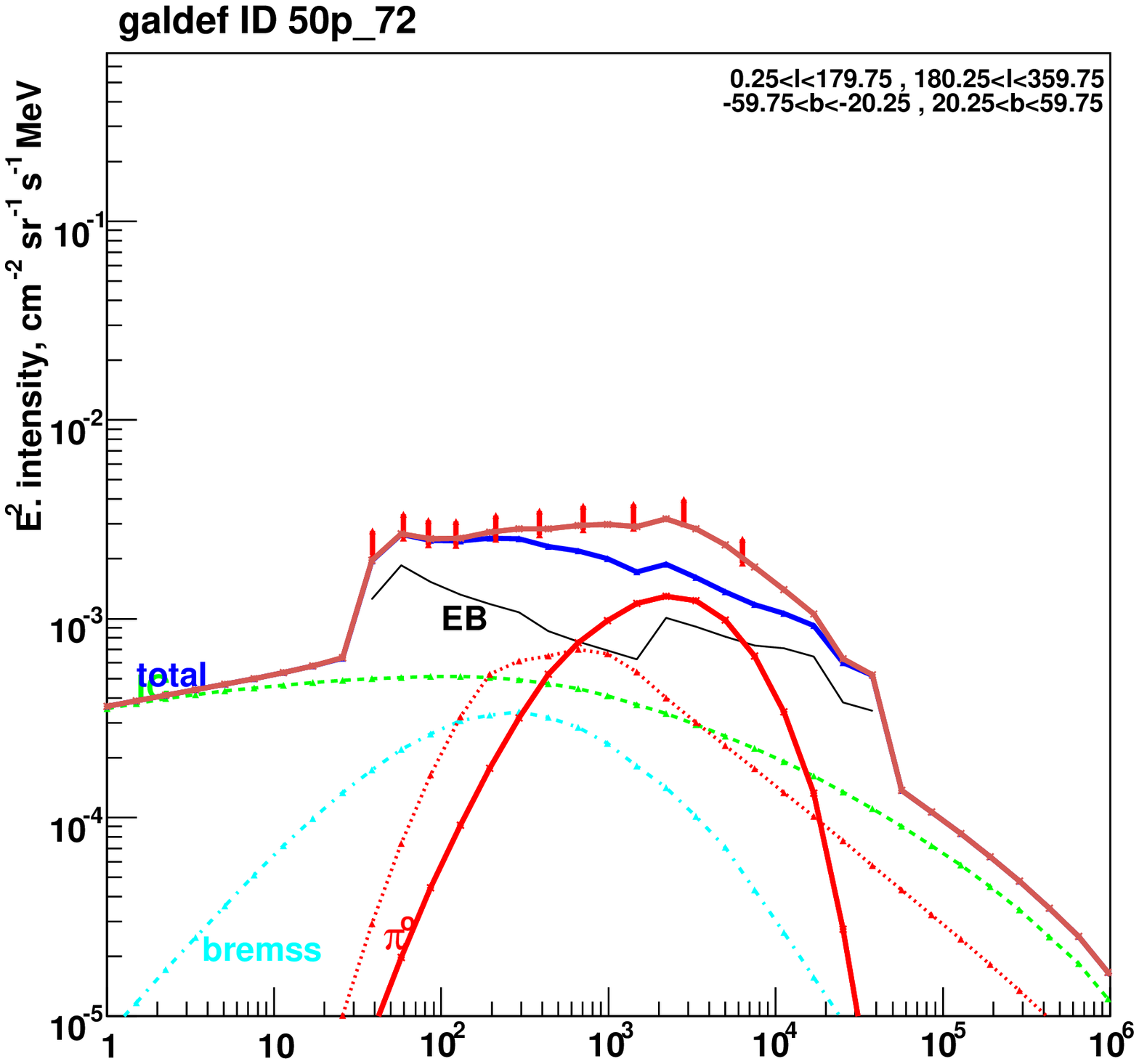}
\includegraphics[scale=0.35]{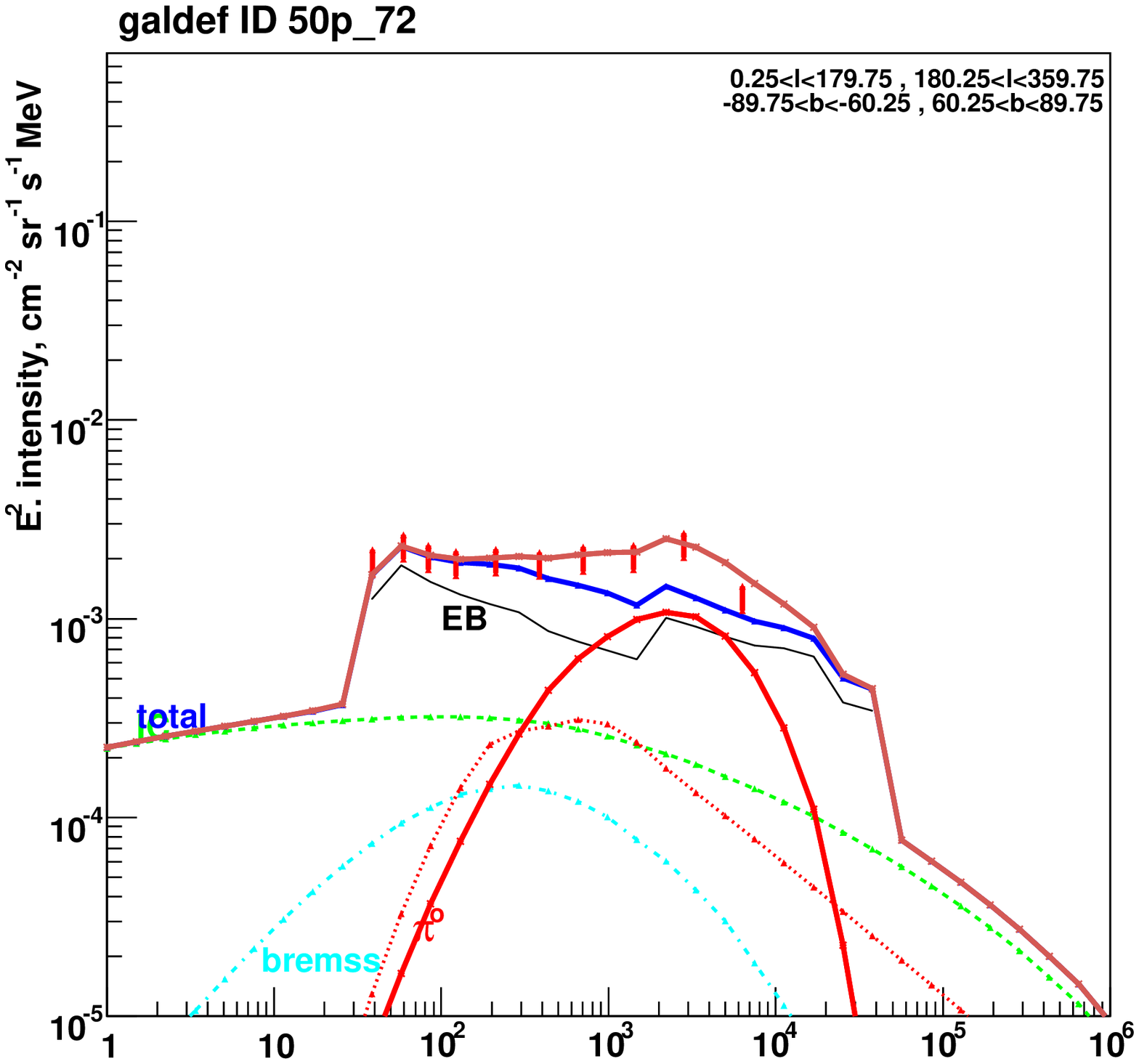}
\caption{\label{result}
Spectra of diffuse $\gamma$-rays for different sky regions
(top row, regions A, B, middle C, D, bottom E, F).
The model components are $\pi^0$ decay, inverse Compton,
bremsstrahlung, EGRB and DMA (dark red curve).
}
\end{figure*}

The results are shown in Fig.~\ref{result}
for six different sky regions as defined in \cite{optimize}.
It should be noted that including the enhancement by subhalos 
dose not exclude the ring-like structures 
proposed by de Boer \cite{boer}.
That is natural since taking the subhalos into account
only enhances the signals coming from the smooth component
but does not mimic the ring-like structure, which can fit
the EGRET data at different directions \cite{boer}. 
Actually the ring-like structure,
 such as the tidal stream of dwarf galaxies are not unusual in N-body
 simulations. Observations and simulations support such an idea that
 the ring at $\sim$14 kpc is from the tidal disruption of the Canis Major
 dwarf galaxy \cite{tidal}.
 Recent result of the rotation curve also predicts a ring like structure
 at the similar position \cite{nrc}. 
From Fig.~\ref{result},
we can see that the EGRET spectra in all regions are
in good agreement with the theoretical values. 
It should also be noted that in our work we adjust the propagation
parameters in GALPROP and do not need an arbitrary normalization of the 
background $\gamma$ rays as done in \cite{boer}. 

\begin{figure}
\includegraphics[scale=0.45]{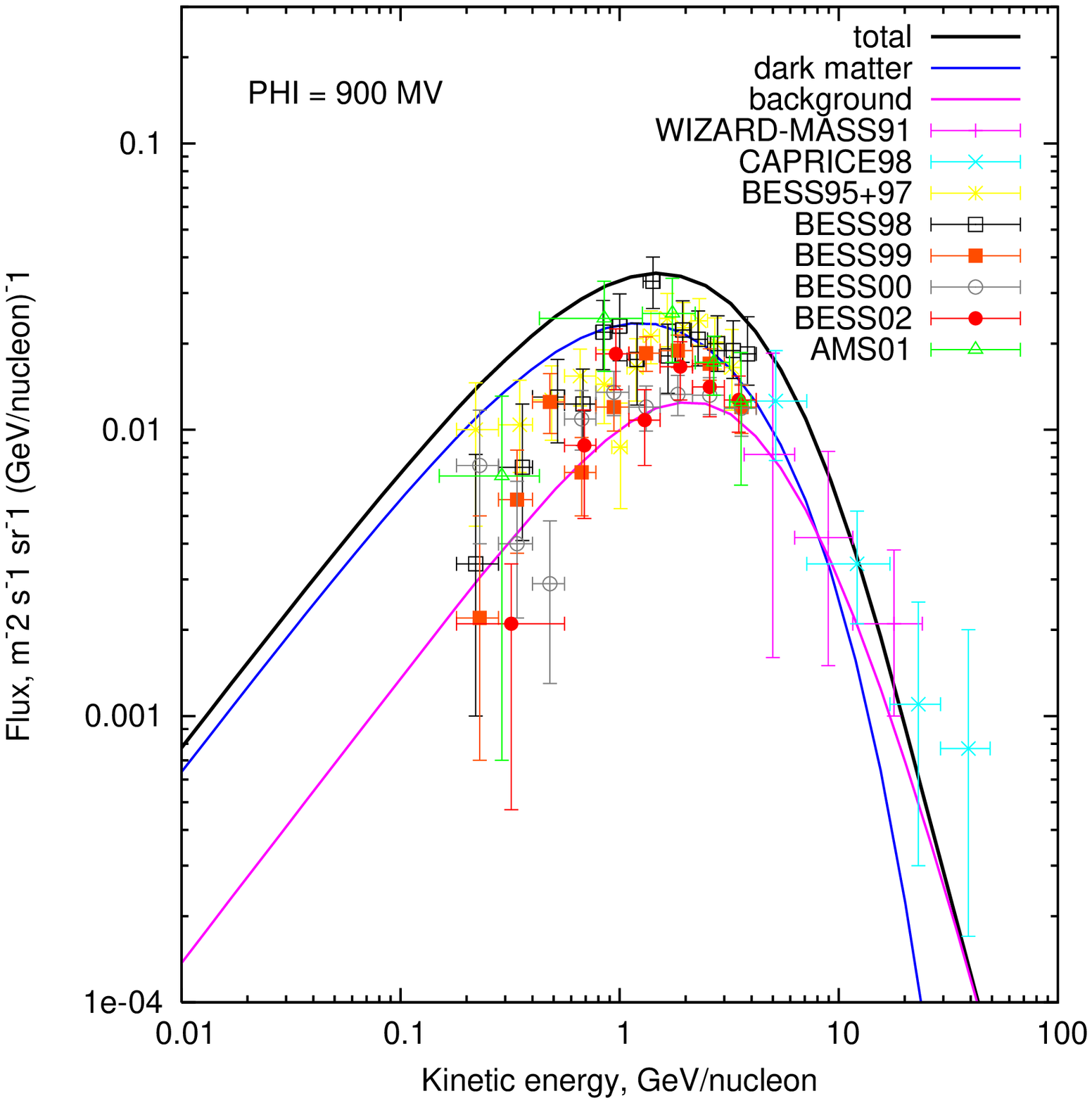}
\caption{\label{pbar}
Flux of ${\bar p}$ after solar modulation.
}
\end{figure}
                                                                              
Finally we check the antiproton flux in this model.
We first calculate the source
term produced by neutralino annihilation adopting the same SUSY model 
as used for $\gamma$-ray calculation, 
$\Phi_{\bar p}(r, E) = \frac{\langle \sigma v \rangle \phi(E)}{2m_\chi^2}
\langle \rho(r)^2 \rangle$
where $\phi(E)$ is the differential flux at energy $E$
by a single annihilation and $\langle \rho(r)^2 \rangle = \rho_{\text smooth}^2
+ \langle \rho_{\text sub}^2 \rangle $. The contribution from the
subhalos is given by $\langle \rho(r)_{\text sub}^2 \rangle =
\int_{m_{min}}^{m_{max}} N_{sub}(m,r)\left( \int \rho^2 
{\text{d}}V \right)\cdot {\text{d}}m $
with $N_{sub}(m,r)$ the number density of subhalos with mass $m$ at radius $r$.
We then calculate the propagation of ${\bar p}$ and its spectrum at 
Earth by incorporating the DMA signals in GALPROP. 
The propagation parameters are kept the same as the ones 
in background $\gamma$-ray calculation.

In Fig.~\ref{pbar} we show the background, signal and total ${\bar p}$
fluxes in our model. The result is much smaller compared with \cite{bergs}. 
Several ways are incorporated to decrease the ${\bar p}$ flux, while keeping   
$\gamma$-rays the same.  
The small $z_h$ in our model helps to suppress the ${\bar p}$ flux from
the smooth DM component. 
The contribution from the rings is found
to be greatly suppressed by slightly adjusting the ring parameters:
the inner ring is now located at R =3.5 kpc and the outer ring is moved
from R = 14 kpc to 16 kpc. This is because the distance dependence of
the propagation ${\bar p}$ is steeper (exponential decrease) than
$r^{-2}$ of $\gamma$-rays \cite{david}. 
It can also be noted that the total ${\bar p}$ flux in the present model
is still a bit higher than the best fit values of the observations at 
lower energies, however, it is consistent with data within $1\sigma$.
The large error of the present data make it hard to give definite conclusion
now. The future measurement from PAMELA \cite{pamela} or AMS02 \cite{ams} 
will finally determine if the present model is confirmed or disproved. 

In summary we calculate the Galactic diffuse $\gamma$-rays from 
CR secondaries and DMA. By building a new
propagation model and taking into account the enhancement
of DMA by subhalos the EGRET data may be explained without any
``boost factor''. However, the ring-like structures are still
necessary. A lower $X_{\text{CO}}$ than previously used value is favored
and the smaller halo height efficiently decrease the yield of ${\bar p}$ from DMA.
The neutralino mass is in the range $40 - 50$~GeV 
and very cuspy profile for subhalos are needed.
The ${\bar p}$ flux coming from secondaries (and tertiaries)
and from DMA are consistent with present experimental bound 
in this propagation model by slightly adjusting the ring parameters.

\begin{acknowledgments}
We thank W. de Boer for helpful discussions on the antiproton flux.
This work is supported by the NSF of China under the grant Nos.
10575111, 10773011 and supported in part by the Chinese Academy 
of Sciences under the grant No. KJCX3-SYW-N2.
\end{acknowledgments}

\end{document}